\documentclass{sig-alternate}
\usepackage{graphicx,url}

\usepackage{verbatim}
\usepackage{color}
\usepackage{float}
\usepackage{multirow}
\usepackage{booktabs}
\usepackage{balance}
\usepackage{array}
\usepackage{pifont}

\newcommand{\textred}[1]{\textcolor{red}{#1}}
\ifx\noeditingmarks\undefined
   \newcommand{\pgwrapper}[2]{\textred{#1 #2}}
\else
   \newcommand{\pgwrapper}[2]{}
\fi

\newcommand*{\twoelementtable}[3][l]%
{%
    \begin{tabular}[t]{@{}#1@{}}%
        #2\tabularnewline
        #3%
    \end{tabular}%
}

\newcommand*\rot{\rotatebox{65}}
\newcommand*\row{\rotatebox{90}}
\newcommand*\OK{{\huge{\ding{51}}}}
\newcommand*\NO{{\huge{\ding{55}}}}
\newcommand*\MA{{\Huge{$\sim$}}}

\begin{document}
%
\conferenceinfo{}{}

\title{Engineering Gesture-Based Authentication Systems\titlenote{Authors' preprint version. To appear in IEEE Pervasive Computing Magazine, Special Issue on Pervasive Security and Privacy, January -- March 2015.}}

\numberofauthors{1} 
\author{
\alignauthor Gradeigh D. Clark and Janne Lindqvist\\
       \affaddr{Rutgers University}\\
       \email{gradeigh.clark@rutgers.edu janne@winlab.rutgers.edu}
}


\maketitle
\begin{abstract}
Gestures are a topic of increasing interest in authentication and successful implementation as a security layer requires reliable gesture recognition. So far much work focuses on new ways to recognize gestures, leaving discussion on the viability of recognition in an authentication scheme to the background.

It is unclear how recognition should be deployed for practical and robust real-world
authentication. In this article, we analyze the effectiveness of different approaches to recognizing
gestures and the potential for use in secure gesture-based 
authentication systems.

\end{abstract}


\section{Introduction}

Authentication has become an essential component in daily life. Increasingly, it is the gateway to critical facets of the human experience such as work, communication, and entertainment. To be effective, any authentication technique must be: reliable, difficult to compromise, and, above all, easy to use when the user is focused on the activity behind the gateway and not the authentication itself.

Gesture-based methods have advantages over current popular authentication methods (e.g.~text entry, PINs, biometrics) given that gestures can be performed faster while also being highly customizable~\cite{mobisys}, easier to remember~\cite{memor}, and potentially more secure~\cite{mobisys}. Gestures require lower concentration and accuracy as compared to other methods, and thus, have potentially lower chances of error when used by stressed or distracted people. For example, an incorrectly entered text-based password yields an automatic rejection whereas some inaccuracy or deviation of the gesture password can still lead to a positive identification.

It is important to note the difference between a \emph{recognizer} and an \emph{authentication system}. The recognizer is one aspect of an authentication system, which can consist of several components (such as the user interface). To the best of our knowledge, there are no comprehensive surveys available of different recognition methods for gestures. Even more importantly, there is no critical discussion available on how these proposals should be compared to each other or how to use them to design robust  and highly usable authentication systems.
Prior studies have looked at basic issues such as asking participants to generate ``secure and memorable'' gesture passwords with no other instructions~\cite{mobisys}. Gesture authentication is at the stage where this work needs to be supplemented by a deeper understanding of usability and effectiveness.
This is not possible until a large open dataset is created that will allow direct comparison of different gesture recognizing methods. Until such a dataset is available, we can qualitatively analyze different approaches to gesture security. Towards this end, our work contributes as follows: 1) survey of common gesture recognizers, 2) 
design considerations for gesture-based authentication systems, 3) suggestions for comparing recognizers, 4) evaluation of gesture-based authentication compared to text-based passwords.
\section{Gesture Types}\label{sec:gesturetypes}

\begin{figure*}
\centering
\includegraphics[scale=0.419]{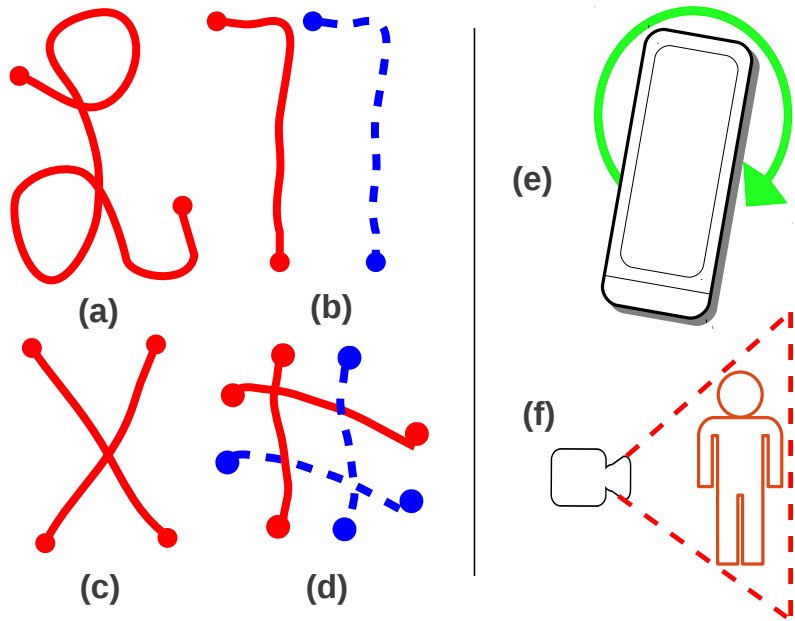}
\caption{An illustration of the different types of gestures. (a) through (d) are different types of touchscreen gestures while (e) and (f) are motion gestures. (a) is a single stroke gesture, (b) is a multitouch gesture, (c) is a multistroke gesture, and (d) is a combination of multistroke and multitouch gestures (the hatch pattern can be drawn by using two fingers and doing two strokes: one stroke with two fingers to get the horizontal portion and a second stroke to form the vertical portion). (e) is an example of a sensor-based motion gesture by rotating a smartphone. (f) is a person being recorded for a camera-based motion gesture. 
}
\label{fig:gestures}
\end{figure*}

There is no universally accepted terminology for gesture types. Often, different names are used for the same type of gesture. From a top-level view, gestures are divisible into two categories: touchscreen gestures and motion gestures. Illustrations of these can be seen in Figure~\ref{fig:gestures}. These two gesture classes can also be free-form (created without constraints and cues) or pre-defined by the creator of a recognizer.

\subsection{Touchscreen} 
Touchscreen gestures are defined as those that are captured through a touchscreen. 
\begin{itemize}
\item Single-stroke gestures use only one finger of the hand to perform a \emph{continuous} input on the screen
\item Multistroke gestures are \emph{discontinuous} gestures that allow for multiple stroke attempts at the screen before completion. 
\item Multitouch gestures use more than one finger to perform a \emph{continuous} gesture. 
\end{itemize}

\subsection{Motion Gestures}
Motion gestures are performed in three dimensions 
and can be divided into sensor-based and camera-based gestures. Sensor-based gestures use other sensors than a camera or touchscreen (e.g.~accelerometer on a smartphone).  This division is motivated by the challenges the input techniques represent for recognizers as well as the abundance of prior work. Camera-based methods represent the majority of publications in gesture recognition, warranting its own category.

\newpage

\section{Threat Models: Attacks Against Gestures}
\label{sec:attacks}

Authentication systems need to be resilient against attacks.
To successfully attack a gesture, an attacker must be capable of replicating the features accurately enough to fool the recognition algorithms into accepting the gesture as authentic. 
There are at least four ways a gesture password can be compromised: shoulder surfing, brute force, dictionary attacks, and storage leakage.

\subsection{Shoulder Surfing}

Shoulder surfing is where an attacker tries to memorize a password or secret via line-of-sight. Some ways to shoulder surf include:
\begin{enumerate}
\item Standard: The attacker observes the user from a vantage point that allows easy viewing of the user's gesture.
\item Recording: The attacker can observe the user's gesture from a recording. 
\item Multiple attackers: Attackers can work together from multiple vantage points to focus on specific parts of a password at different times and reconstruct it later.
\end{enumerate}

\subsection{Brute Force}
Brute force attacking is done by repeatedly trying passwords to find the right one. This can be a way to measure the susceptibility of a recognizer to algorithmic attack.
An equivalent attack on text-based passwords would be trying to guess the password without having an access to the hashes of the passwords (see: Storage Leakage). 

\subsection{Dictionary Attack}
A dictionary attack is similar to a brute force attack except that the password attempts come from a set of more likely possibilities (such as datasets from user studies).
Dictionary attacks have been demonstrated against gestures. A robot was built to perform gesture inputs on a smartphone based on training data from human participants~\cite{Serwadda:2013:KTB:2508859.2516659}. The robot was capable of successfully authenticating for at least half of the gestures in the sample population. However, this approach relied on a simplified assumption that the touchscreen device would not lock itself after numerous unsuccessful attempts.

\subsection{Storage Leakage}
Storage leakage can be a problem depending on how the gesture password is stored on the device. To successfully steal a gesture based on stored data, a thief would have to know both the structure of the recognizer and how to translate the stored values into gesture actions. Text-based passwords mitigate storage leakage by only storing the cryptographic hash of the password.
Storage leakage is a serious issue for gesture passwords given that no two inputs will be exactly alike, which makes comparing their hashes difficult.
\section{Comparing Authentication Systems}

Before discussing recognition in full, it is useful to outline aspects of gestures as an authentication scheme to motivate their viability as a replacement for (or supplement to) text-based passwords. The best way to do that is to evaluate gestures on three criteria: usability, deployability, and security. These three metrics follow from an exhaustive survey of authentication methods~\cite{bonneau}, which unfortunately does not discuss gestures. Usability addresses the viability of a scheme from the perspective of the user. Deployability examines what infrastructure a method requires in order to be usable. Security refers to the ability of a system to resist attacks. 

All known alternative systems do not have the full range of features offered by text-based passwords~\cite{bonneau} and gestures are no exception to this. Intentionally or not, decades of infrastructure and development has gone into making text-based passwords ubiquitous. Because of this, and in spite of the many disadvantages of text-based passwords, it is difficult for any scheme to match all the benefits they have. Advancing the development of alternative schemes could allow for a real challenger to text-based passwords.

\subsection{Usability}

Gestures are potentially more memorable than text-based passwords because human recall is better for pictorial concepts rather than strings of text~\cite{memor}, although there is no definitive measure of the memorability of a password.
It cannot be yet stated whether complicated gesture passwords are more memorable than complicated text-based ones, but there is evidence from human psychology~\cite{memor} and user studies~\cite{mobisys} to support the assertion.

Gesture-based passwords are as easy to adopt as text-based passwords. Most people have used gestures as a way to communicate silently or even drawn pictures to explain something to another person. Little additional training is required to teach people how to use them. The introduction of the Android 3x3 grid-based graphical password can be thought of as a primer to using gesture-based passwords on touchscreen devices.

Latency and error rates are natural usability related concerns, and can reduced with proper recognizer design. Error rates are harder to minimize. Complicated gesture passwords \emph{could} have better rates than complicated text-based passwords. 

Password recovery or the ability to reset your password is necessary for usability. Gestures are equal to text-based passwords in this -- the same systems in place that exist to recover text-based passwords can be applied to gesture passwords (see Deployability). The password recovery flow for gestures will be different than with text-based passwords since the features of a gesture can be used to recover their password. A simple example would be asking a user to trace a set of characters and seeing how that correlates to their past behavior. Proper users can either be shown a picture of their gesture or given steps or hints as to how it can be replicated.

The effort required to authenticate depends on the population under consideration. Given the ubiquity of text-based passwords, user effort \emph{can} be higher when starting out with gestures since the users are primed to more comfortably use keyboards. On the other hand, Android users could be primed by the 3x3 graphical password system towards touchscreen gestures. However, differently-abled users  (e.g.~paraplegics) who cannot naturally interact with a smartphone, touchscreen, or properly motion to a camera are at a disadvantage.

\subsection{Deployability}

Gesture passwords, with current technology, cannot be used by all the people who can use text-based passwords. The differently-abled can face issues using gestures, especially if they suffer from a loss of sight or motor function.

Users do not require additional tools to authenticate.  This does not mean that gestures are compatible with all current systems -- rather that they integrate well into existing infrastructures. Although touchscreen tablets and phones are already prominent, laptops and monitors with touch capability are starting to become more commonplace as well. Alternatively, gestures could be generated with a mouse or using the touch-interactive mousepads on laptops. A negative is that, for motion gestures, desktop users may require a separate webcam component.

Gestures are not server compatible -- this is attributable to both their infancy as an authentication scheme and the proliferation of text-based passwords. Some devices use biometrics as a master password, allowing integration with servers. A gesture could be used as a master password in much the same way. This would improve server compatibility, but it is not a perfect solution -- the problem stems from the inability to reliably compare the hash of two different gesture inputs. 

Gestures can be integrated into web browsers. HTML5 or mobile websites can have a gesture capture area for touchscreen and allowing a browser access to the camera would enable camera-based gestures. A mobile platform could observe a motion gesture on behalf of the browser. Using gestures may require additional hardware on desktops (e.g.~webcam, external sensors), though these are becoming default equipment for desktops, too.

\subsection{Security}

Gestures can be more resistant to shoulder surfing attacks than text-based passwords, depending on the amount of features used in recognition. Replication of the exact way a gesture is performed can be more difficult than assembling all the characters of a text-based password, however, this depends on the password length. Similar to biometric systems, personal knowledge does not yield clues that could reveal a user's gesture password. Comparing the security of text-based passwords to gesture passwords is an open problem. It is possible to quantify the security of a gesture password based on a ``surprisingness'' factor~\cite{mobisys}. This score allows for password creation policies similar to text-based passwords (e.g.~rejecting ``simple'' passwords and instructing the user to try again).

If a system restricts the number of failed attempts, then it becomes difficult to compromise the password. As with text-based passwords, gesture passwords can be stolen if the attempts are not limited and if the attacker is properly trained~\cite{Serwadda:2013:KTB:2508859.2516659}. The lower accuracy needed for gestures, an advantage in usability, is a disadvantage here.  This demonstrates the need for proper gesture password creation policies such as the ones used for text-based passwords.

Storage leakage is a problem because of the fact that there is not yet a way to store gestures such that two similar inputs would have the same cryptographic hash. One way could be to train a recognizer to quantize the data within very small intervals. The intervals should be chosen so that users could reliably perform inside them. Quantized data can be appended to a string and then hashed. The downside is that this approach could have large error rates and would require more precision when performing the password than what would normally be required. 

\section{Designing Recognizers for Authentication}

Some design considerations can be learned from prior work on recognizers~\cite{Wobbrock:2007:GWL:1294211.1294238}, although the focus there was on recognition and not authentication. Additionally, some of the aspects for reliable recognition~\cite{Wobbrock:2007:GWL:1294211.1294238} do not translate to reliable authentication (e.g.~location and scale invariance). We extend the previous work with design considerations for authentication systems. 

\begin{enumerate}
\item Sample invariance. Different devices have different sampling rates. Additionally, there are natural variances in the speed and time with which a user performs gestures between authentication attempts. 
 A recognizer should resample the input to obtain an accurate portrait of the gesture while keeping the number of samples constant.

\item Trainable. A good recognizer allows for the design and learning to handle new inputs. It should not use only predefined gestures. It should also differentiate between similar gestures (e.g.~drawing a rectangle versus a square). Users need to be allowed to create their own gestures to maximize usability and comfort while leveraging the full utility of the password space.

\item Adaptive. User behavior can change over time. For example, users will learn to perform their gesture password with higher speeds. The recognizer should adapt to these changes over time. It should work through stored templates and features long after the initial training phase to figure out which templates are working and which ones are not.
\item Computationally efficient. When designing efficient recognizers, it is necessary to minimize the overall computation, memory, and delay introduced by the algorithm. The overall user experience is degraded if there is a noticeable pause with every login attempt.

\item Storage conscious. Recognizers should not make the system unusable by storing a large number of templates or extracted features. The gestures should also be protected from theft by straightforward copying.

\item Configurable. A gesture recognizer should allow options for users and developers. Examples include control over the sampling rate or how many stored templates to use.  Users could also configure security settings based on personal needs. 

\item Attack resistant. A recognizer must be efficient at rejecting false users. Gestures are represented as a collection of features. These features form layers that increase resistance to attacks. With more features, a recognizer increases its ability to exclude impostors. Examples of these features are: pressure, speed, finger or arm length, body type, and path length. Recognizers with more than one layer can be considered attack resistant.

\item API-friendly. Although recognizers can be described in papers and with pseudocode, it may not be enough to be understandable to developers. A difficult, non-intuitive recognizer can have adoption issues. Developers that benefit from it (e.g.~college students or startups) would have trouble using the recognizer if it is too difficult to implement.
\end{enumerate}

\section{Survey of Common Approaches to Gesture Recognition}

A gesture recognizer uses algorithms to interpret human gestures.

Designers have been creating new recognizers for either a neglected gesture type or for supporting new features. This has led to numerous innovations in recognition for different gesture types. However, from a security engineering perspective, there is not enough consideration put towards authentication. Below, we analyze several popular algorithms and in Table~\ref{fig:sum} we summarize how these algorithms correlate to the design principles of the previous section.

\subsection{Geometric Methods}
This family of recognizers performs distance-based comparisons on stored templates of coordinate pairs. 
For touchscreen gestures, the comparative measure is coordinate pairs in the plane. For motion gestures, the measure is either accelerometer or gyroscope data side of a volume~\cite{pro3d}. The following discussion applies to several recognizers including \$1~\cite{Wobbrock:2007:GWL:1294211.1294238}, \$N~\cite{Anthony:2010:LMR:1839214.1839258} and Protractor~\cite{Li:2010:PFA:1753326.1753654} for touchscreens. Protractor3D~\cite{pro3d} is an extension for motion gestures. All of these recognizers perform at least these first four out of five steps:
\begin{enumerate}
\item The gesture is resampled to $N$ points.
\item The resampled gesture is translated to the origin.
\item The size is normalized so the points are contained within a bounded cube.
\item  It is then rotated until the angle that the first point in the sequence makes with the centroid of the gesture sequence is zero.
\item The gesture is then iteratively rotated until an alignment is found that produces the optimal score with a given template.
\end{enumerate}

A motion gesture recognizer does not need to consider rotation, and therefore the fifth step above does not apply to it. The concern is with the difference between successive accelerometer and gyroscope readings. 

Geometric recognizers are sample-invariant and can be trained to identify new gestures. They are partially adaptable since they could store every single successful attempt as a new template at the expense of both storage and efficient computation, although this renders early templates meaningless. The algorithms are computationally efficient since templates are stored in a pre-processed form, requiring only the new attempt to be processed. Since they store little more than geometric features, but do not protect the data, they are partially storage conscious. They are configurable since a user can be granted control over every step in the algorithm. Geometric recognizers are partially attack resistant since they only have one layer of features to breach. Finally, they are API-friendly since they perform simple operations on coordinate pairs.

\subsection{Dynamic Time Warping}
Successive inputs tend to be mismatched because it is difficult for users to enter their gestures the exact same way every time. For example, a user might have rounded a corner more sharply in one attempt or the other input is drawn more slowly and carefully. From a top-level view, these attempts will often appear identical. Where they differ is in the details -- one will have more sampling points than the other. The distribution of the inputs, when plotted against time, will often be very similar to each other -- the difference being that they may \emph{appear} as time shifted versions of each other. Dynamic Time Warping (DTW) transforms the gestures such that they can be compared directly.

DTW fixes this misalignment by stretching the gestures such that they are aligned perfectly in time -- the way this is done is by repeating sampling points in areas where one input is shorter than the other. So an $N$x$M$ matrix of path differences between gestures is constructed and traversed in a way that resamples each gesture to an equal length.  After that,  point-wise distance comparisons can be made and measured against a threshold to determine authentication~\cite{Sae-Bae:2012:BGN:2207676.2208543,kinwrite}.

DTW is sample invariant due to the alignment of time series. Distance based measures are used to do recognition on stored templates after alignment, allowing for trainability. It is partially adaptable, using the same logic applied with the geometric methods above. It is partially configurable, since the only aspect the user can control is the authentication threshold and the template count.  Common DTW implementation is computationally inefficient, but there exist implementations that can reduce the computation time. However, that results in tradeoffs in recognition efficiency, leading to it being classified as partially computationally efficient. DTW is partially storage conscious, requiring no extra data than a small number of templates but it makes no provisions on protecting the data. It is partially attack resistant given that it is based on one layer of resistance (coordinate pairs). It is partially API-friendly, because it is not straightforward to implement path alignment and to reduce the complexity beyond the general implementation.

\begin{table*} \centering
    \begin{tabular}{@{} cl*{10}c @{}}
        & & \multicolumn{10}{c}{\large \textbf{\underline{Features}}} \\[2ex]
        & & \rot{{\shortstack[l]{Sample \\ Invariant}}} & \rot{Trainable} & \rot{Adaptive} & \rot{{\shortstack[l]{Computationally \\ Efficient}}} & \rot{{\shortstack[l]{Storage\\Conscious}}} 
        & \rot{Configurable} & \rot{{\shortstack[l]{Attack \\ Resistant}}}
        & \rot{API-Friendly}  \\
        \cmidrule{2-12}
        & Geometric             	& \OK & \OK  & \MA &\OK  &\MA   &\OK  & \MA & \OK  &   & \\
        & DTW	                   	& \OK & \OK & \MA & \MA & \MA & \MA & \MA & \MA &  & \\
        & HMM 			& \OK & \OK & \OK & \MA & \NO & \OK & \OK & \NO   &  & \\
        & SVM			& \OK & \OK & \OK & \MA & \MA &  \OK & \OK & \NO &  & \\
        & WiFi		& \NO & \MA & \NO & \NO & \OK &  \NO & \NO & \MA &  &  \\
        & Capacitive		& \NO & \MA & \NO & \OK & \OK &  \NO & \NO & \NO &  &  \\
 \row{\rlap{\large{ \textbf{\underline{~Recognizers}}}}}
                & RFID			& \NO & \MA & \NO & \OK & \OK &  \NO & \NO & \NO &  &  \\
        \cmidrule[1pt]{2-12}
    \end{tabular}
    \caption{A summary of the design considerations as applied to the described recognizers. See each section for justification on the rating system. 
    }
    Legend:\\
    \OK = Has the feature\\
    \MA = Has aspects of the feature, but not fully realized\\
    \NO = Does not offer the feature\\
    \label{fig:sum}
\end{table*}

\subsection{Machine Learning Techniques}
Machine learning studies algorithms to teach computer how to perform tasks from data. This is a natural fit for performing gesture recognition.  
The most popular methods are Hidden Markov Models (HMMs) and Support Vector Machines (SVMs). 
\subsubsection{Hidden Markov Models}

For HMM recognizers, the user's input appears as a collection of features (e.g.~time, pressure, distance) and not as a known gesture. The sequence of states (the Markov process) a gesture undergoes (e.g.~up, left) cannot be seen, only the measurable outcomes (it lasted for $t$ time and has $N$ samples).

An HMM consists of a set of states, a transition probability matrix describing the chance to transition from one state to another, and an output probability function. Each individual gesture to be recognized for a system requires its own, pre-specified HMM. Given the input data, the recognition task is to figure out a sequence of state transitions a user's gesture might be mapped to. There is more than one possible state sequence for any input. The most likely sequence (and thus gesture) is determined by evaluating the joint probability of the sequence and observations~\cite{rabiner}.

\subsubsection{Support Vector Machines}
SVMs are a set of supervised learning algorithms that can be used for classification and recognition problems~\cite{vapnik}.
SVMs solve a binary classification problem. In the context of gestures, the two classes would be a specific gesture (Class 1) and ``not a gesture'' (Class 2). A collection of gestures can be split into one of these classes depending on what is being recognized. If plotted in a plane, there should be a visual way of drawing a boundary such that Class 1 and Class 2 data points are separated from each other. The Class 1 and Class 2 data points that are closest to each other (and thus closest to the boundary) are called the support vectors. These are the most difficult data points to classify since they most closely affect the location and contour of the boundary.

SVMs are useful because support a large number of variables, representing them as vectors of features. However, the data will need to be mapped to a higher dimensional space proportional to the number of features. In order to create a decision boundary,  a non-linear equation must be solved. The training data is used to construct the boundary and new user input is classified as a gesture depending on what side of the boundary it ends up on.

Both HMM and SVM do not depend on the number of samples; more is often better. Both can be trained to accept any input. Since learning is a continuous process, both can be made to adapt over long periods to user input. Computationally, they are the slowest algorithms presented and are partially computationally efficient because of this.  HMMs require a large number of training examples due to the many different paths a gesture can take, requiring more space. SVMs can function properly with a low number of examples, depending on the basis set and complexity of the recognition space. Neither HMMs nor SVMs store training data in a way that protects it from leakage. As such, HMMs are not storage conscious and SVMs are only partially conscious. HMMs and SVMs have more layers of resistance to attacks since they are capable of using many different features (not just coordinate and time data) to recognize gestures. They are, however, difficult to program and are not API-friendly.

\subsection{Other Gesture Recognition Methods}

 A gesture could be identified by measuring very small, finite Doppler shifts between incoming and outgoing WiFi signals~\cite{puetal}. If a user is wearing an RFID tag, an array of antennas can be used to track the disturbance caused by a gesture~\cite{rfid}. A tag moving between the antennas generates readings which are fed back wirelessly and analyzed to determine what gesture is being performed. Similarly, there are demonstrated methods for hand gesture recognition with capacitive proximity sensors. A person waving their hand would generate readings of different magnitudes at different timestamps, from which the gesture can be determined~\cite{cap}.

These methods are not sample invariant -- their recognition method is based on discovering patterns in data, which sampling to a constant can affect. All three have some degree of trainability since they are capable of learning some gestures. However, there can be resolution issues between recognizing two similar gestures with these methods that is difficult to resolve. There are no provisions for adaptivity. While capacitive and RFID recognition are computationally feasible, WiFi recognition has involved steps due to various transforms needed to analyze the input data so it is not computationally feasible. All three are neither configurable nor resistant; they have difficulty distinguishing users and focus only on gestures. Finally, only WiFi and RFID are partially API-friendly since common equipment are available (e.g.~routers, tags), proximity sensors require special equipment and programming.

\section{Cross Recognizer Comparisons}

It is useful to understand how different recognizers reliably compare to each other. Typical measures of performance used for recognition are: Receiver Operating Characteristic (ROC) curves and Equivalent Error Rates (EERs).

An ROC curve is a plot of a user's successful login rate to a system versus an attacker's successful login rate as the threshold to authenticate is varied from zero (accept any input as the password) to infinity (reject any input, even the real password). As the threshold increases, the rate of false positives (FPR) decreases and the rate of true positives (TPR) increases (up until a certain point, at which it then decreases). Varying this threshold generates a new (TPR,FPR) point at every step -- the plot of these together forms the ROC curve. The EER is the rate at which the number of accepted attackers and rejected true users are equal. Perfect recognizers would have a TPR of 100\% (all true user attempts accepted), an FPR of 0\% (all attacker attempts rejected), and an EER of 0\% (0\% rejected true users and 0\% accepted attackers).

Ideally, the recognizer with the lowest reported EER value would be the best one to use. But that intuition fails since recognizers are rarely, if ever, computed across same datasets; most designers generate new datasets to evaluate their system. In an authentication system, just recognizing a gesture is not enough, we need to know whose gesture it is. 

Many recognizers focus on differentiating a narrow vocabulary of gestures (e.g.~a circle from a rectangle) and report results based on that. 

For authentication, it is necessary to compute EER values based on attacks against a gesture. Most recognizer designers do not do this. To justify the validity of recognition techniques, there needs to be a common reference point to compare the EERs.

A large and public dataset of gestures that are meant for testing recognizers is a necessary step for comprehensive recognizer comparison. This is not a unique problem. Optical Character Recognition (OCR) algorithms faced the exact problem that gesture recognizers face now. There was no way to adequately compare all of the different methods until the U.S.~Department of Energy commissioned the large scale creation of a comprehensive OCR dataset that developers could use to test their algorithms. 
A similar situation was present in speech recognition problems as well. The unifying theme here is that recognition algorithms are \emph{not} directly comparable without an extensive public dataset.

Many recognition algorithms are dependent on the input data type, complicating the process of creating a dataset. However, there can be issues if not enough features are gathered from a gesture. As an example, a dataset compiled around motion gestures on a smartphone may collect time, coordinate, accelerometer, and gyroscope data. But there could be a new recognizer that might need data from the gravity or magnetometer sensors. The dataset should be left entirely open so new features can be added when needed.

The sets used for benchmarking should be categorized based on specific test criteria. Example sets include: weak passwords,  strong passwords, memorable passwords, and memorable \emph{and} strong passwords.  The gesture type is also an important variable.

Since there is no public dataset yet, comparing the recognizers for authentication purposes, requires answering the following questions:
\begin{enumerate}
\item What is the dataset that the recognizers are being compared to?
\item How do these recognizers compare to one another based on the design criteria (Table 1)?
\item Were the EER values computed for the algorithm subject to gesture-based attacks?
\item How do error rates vary as features are added or subtracted?
\end{enumerate}

\section{Conclusions}

We divided gestures into two categories: touchscreen and motion-based gestures. The attacks against gestures include common methods to attack passwords such as brute force, dictionary attacks, and storage leakage. We used the usability, deployability, and security framework~\cite{bonneau} to highlight areas where gestures are not as good as text-based passwords. Specifically: they are not usable by members of the differently-abled population, they do not yet integrate seamlessly into Internet architectures (making them more suitable for personal devices and offline security solutions), and that they are susceptible to storage leakage with no reliable way to compare hashes of two attempts.

We discussed design considerations for gesture recognizers: sample invariance, trainability, adaptivity, computational efficiency, storage consciousness, configurability, attack resistance, and API-friendliness. Of the most common gesture recognition algorithms, none appear to offer all features simultaneously. SVMs currently satisfy most of the considerations. Many recognizers that exist are not inherently designed to authenticate, but rather to recognize distinct gestures.

The presented algorithms are not directly comparable on the basis of EER since none of them are computed using the same data. Gesture recognition requires a large public dataset to allow comparison, similar to optical character recognition and speech recognition.

Finally, gesture-based authentication systems show potential for practical adoption. Although they have disadvantages, these can be attributed to the lack of development in recognizers and infrastructure. Text-based passwords work well for users that are comfortable with physical keyboards, yet touchscreen keyboards are becoming more and more common with the proliferation of mobile devices. Eventually, touchscreen keyboards may replace physical keyboards altogether. On those interfaces, gestures become a more natural authentication choice than a text-based password because it is more difficult to type character strings. However, the ubiquity of text-based passwords makes it exceedingly difficult for alternative methods to gain traction. The potential benefits of gestures should not be ignored because of the current limitations.

\section{Acknowledgments}
This material is based upon work supported by the National Science
Foundation under Grant Numbers 1223977 and 1228777. Any opinions, findings,
and conclusions or recommendations expressed in this material are
those of the authors and do not necessarily reflect the views of the
National Science Foundation.

\balance
\bibliographystyle{abbrv}
\bibliography{gestures}  
%
%
\balancecolumns

\end{document}